\documentclass[aps,preprint,superscriptaddress,a4paper]{revtex4-2}

\usepackage{graphicx}

\usepackage{mathtools}
\usepackage[colorlinks = true,linkcolor = blue,urlcolor  = blue,citecolor = blue,anchorcolor = blue]{hyperref}
\usepackage{xcolor, soul}
\usepackage{color}

\usepackage[draft]{changes}

\DeclarePairedDelimiter\ket{\lvert}{\rangle}

\def\phalf{+\frac{1}{2}}
\def\mhalf{-\frac{1}{2}}
\def\pthalf{+\frac{3}{2}}
\def\mthalf{-\frac{3}{2}}

\begin{document}

\title{Demonstrating experimentally the encoding and dynamics of an error-correctable logical qubit on a hyperfine-coupled nuclear spin qudit }

\author{Sumin Lim}
\affiliation{CAESR, Department of Physics, University of Oxford, The Clarendon Laboratory, Parks Road, Oxford OX1 3PU, UK}

\author{Mikhail V.\ Vaganov}
\affiliation{CAESR, Department of Physics, University of Oxford, The Clarendon Laboratory, Parks Road, Oxford OX1 3PU, UK}

\author{Junjie Liu}
\affiliation{CAESR, Department of Physics, University of Oxford, The Clarendon Laboratory, Parks Road, Oxford OX1 3PU, UK}
\affiliation{School of Physical and Chemical Sciences, Queen Mary University of London, London E1 4NS, UK}

\author{Arzhang Ardavan}
\affiliation{CAESR, Department of Physics, University of Oxford, The Clarendon Laboratory, Parks Road, Oxford OX1 3PU, UK}

\maketitle

{Abstract:  
The realization of effective quantum error correction protocols remains a central challenge in the development of scalable quantum computers. Employing high-dimensional quantum systems (qudits) can offer more hardware-efficient protocols than qubit-based approaches. Using electron-nuclear double resonance, we implement a logical qubit encoded on the four states of a {\sl I}~=~3/2 nuclear spin hyperfine-coupled to a {\sl S}~=~1/2 electron spin qubit; the encoding protects against the dominant decoherence mechanism in such systems, fluctuations of the quantizing magnetic field. We explore the dynamics of the encoded state both under a controlled application of the fluctuation and under natural decoherence processes. Our results confirm the potential of these proposals for practical, implementable, fault tolerant quantum memories.}

While the control fidelity in few-qubit systems has reached two or three nines \cite{noiri2022fast, xue2022quantum, yoneda2018quantum, yang2019silicon}, maintaining this becomes increasingly difficult as the size of the system is scaled up to tens of qubits~\cite{preskill2021quantum, preskill2018quantum}. Regardless of the physical platform, unwanted inter-qubit interactions and coupling to the environment drastically reduce the overall fidelity of the quantum computing system. 
Quantum error correction protocols seek to address these challenges by enlarging the Hilbert space to store information with redundency~\cite{terhal2015quantum, gottesman1997stabilizer}. Many current proposals are based on the idea of using multiple physical qubits to encode an individual logical qubit~\cite{divincenzo1996fault, laflamme1996perfect, knill1997theory, knill2000theory}, yielding qubit-based error correction algorithms such as surface codes~\cite{dennis2002topological, chen2021exponential}. An alternative paradigm deploys arbitrary $d$-dimensional quantum systems (where $ d > 2 $), known as qudits. This approach reduces the overhead associated with additional qubits, and can provide a hardware-efficient structure. Originally proposed by Gottesman, Kitaev, and Preskill~\cite{gottesman2001encoding}, a range of theoretical~\cite{pirandola2008minimal, noh2020encoding, albert2020robust, royer2020stabilization} and experimental~\cite{hu2019quantum, fluhmann2020direct, ofek2016extending, campagne2020quantum, heeres2017implementing, gertler2021protecting} explorations are ongoing into systems based on quantum harmonic oscillators with bosonic eigenstates. This kind of approach provides infinite dimensions in a spatially compact system, but at the cost of complexity in practical operations. 

In this context, quantum spins greater than 1/2 offer a promising alternative: a single physical object providing a multi-dimensional but finite and, often, well isolated Hilbert space. Electronic spins in condensed matter, with their well-defined quantum properties and relatively weak interactions with external excitations, are natural candidates for embodying quantum information; their hyperfine-coupled nuclear spins, which tend to be even more coherent, offer potential as quantum memory elements. Indeed, such systems featured among the earliest theoretical condensed matter quantum information proposals such as that of Kane~\cite{kane1998silicon}, and various physical examples have been studied experimentally, including paramagnetic defects in semiconductors~\cite{morton2008solid, Morello2010-ff,Pla2013-rk} and diamond~\cite{Waldherr2014-oo,taminiau2014universal, robledo2011high, bradley2019ten} 
and molecular magnets~\cite{Vincent2012-mi,Thiele2014-il}. Together, these observations motivate us to explore how a nuclear spin qudit hyperfine-coupled to an electron spin qubit~\cite{asaad2020coherent,  Fernandez2024, Yu2024-bj, chiesa2021embedded, gross2024hardware, chicco2023proof}
can be used to encode fault-tolerant error correcting protocols~\cite{Gross2021-qx, chiesa2020molecular, Lim2023-yy}.

A particularly simple proposal employs an electron spin $S=1/2$ hyperfine-coupled to a nuclear spin $I=3/2$ to encode logical qubit states protected against fluctuations of the magnetic field $\Delta B_z$ along the quantisation axis, represented by 
\begin{equation}
\hat{O}_\mathrm{fluc.} = (\gamma_\mathrm{S} \hat{S_z} + \gamma_\mathrm{I} \hat{I_z})\Delta B_z
\end{equation}
where $\gamma_\mathrm{S}$ and $\gamma_\mathrm{I}$ are the electron and nuclear spin gyromagnetic ratios respectively.
A qubit state $\ket{\psi}$ encoded on the electron spin with a particular projection of the nuclear spin can be represented in the notation $\ket{m_S}\otimes\ket{m_I}$ as
\begin{equation}
\ket{\psi} = \left(\alpha\ket*{\mhalf} + \beta 
\ket*{\phalf}\right)
\otimes \ket*{\mhalf} 
\label{eqn-electron-qubit}
\end{equation} 
where $\alpha$ and $\beta$ are, in general, complex.
This state is rotated to the protected ``encoded'' state
\begin{equation}
\ket{\psi_L} = \alpha\ket{0_L} + \beta 
\ket{1_L}
\label{eqn-encoded-state}
\end{equation}
where 
\begin{eqnarray}
\ket{0_L}=\ket*{\mhalf} \otimes 
\left( \frac{1}{2}\ket*{\mthalf}+\frac{\sqrt{3}}{2}\ket*{\phalf} \right), \nonumber \\ 
\ket{1_L}=\ket*{\mhalf} \otimes 
\left( \frac{1}{2}\ket*{\pthalf}+\frac{\sqrt{3}}{2}\ket*{\mhalf} \right) \; .
\label{eqn-encoding}
\end{eqnarray}
The logical states $\ket{0_L}$ and $\ket{1_L}$ are chosen such that they and the states connected by the perturbation against which we wish to protect (i.e., $\hat{O}_\mathrm{fluc.}\ket{0_L}$ and $\hat{O}_\mathrm{fluc.}\ket{1_L}$) form a mutually orthogonal set of states. Note that the electron spin part of $\ket{0_L}$ and $\ket{1_L}$ are eigenstates of $\hat{S_z}$ with the same $m_S$; because $\gamma_\mathrm{S}$ is generally larger than $\gamma_\mathrm{I}$, encoding the qubit in the nuclear degrees of freedom already provides an element of protection against magnetic field fluctuations~\cite{morton2008solid}. The result of the magnetic field fluctuation acting for time $\delta t$ on the logical qubit state is therefore
\begin{eqnarray}
\exp \left( -\frac{\mathrm{i}}{\hbar} \hat{O}_\mathrm{fluc.} \delta t \right) \ket{\psi_L}  \nonumber \\
= \exp \left( \mathrm{i} \frac{\omega_S}{2} \delta t \right) \cdot
\exp{\left(-\mathrm{i} \theta \hat{I_z} \right) \ket{\psi_L}}
\label{eqn-exact-errored-state}
\end{eqnarray}
where $\omega_S = \gamma_S \Delta B_z$ and $\theta = \gamma_I \Delta B_z \delta t / \hbar$. The first factor is an overall phase which is undetectable for this system in isolation; we shall neglect it in what follows. Expanding the second factor,
\begin{eqnarray}
\exp{\left(-\mathrm{i} \theta \hat{I_z} \right) \ket{\psi_L}} & = &
\sum_n A_n \frac{(-\mathrm{i}\theta)^n}{n!} \hat{I_z}^n \ket{\psi_L} \nonumber \\
& \approx & A_0 \ket{\psi_L} -  \mathrm{i}\, A_1\, \theta\, \hat{I_z} \ket{\psi_L} 
\label{eqn-errored-state}
\end{eqnarray}
where the approximation holds for a weak perturbation, i.e.\ a small $\Delta B_z$ acting for a short time $\delta t$. 
 The effect of the perturbation is to shift amplitude from the logical qubit subspace into the (by construction) orthogonal ``error'' subspace spanned by $\hat{I_z}\ket{0_L}$ and $\hat{I_z}\ket{1_L}$. Here $A_n = 1$ for all $n$; we shall later use $A_n$ to parameterise our experimental results.  

The fact that the uncorrupted component (the first term in Eqn.~\ref{eqn-errored-state}) is orthogonal to the corrupted component (the second term) allows us to identify a unitary transformation (see Supplementary Information Section~I) ``decoding''  \cite{chiesa2020molecular} this state to
\begin{eqnarray}
A_0 \ket*{\mhalf} \otimes \left( \alpha \ket*{\mhalf} + \beta \ket*{\phalf} \right) & \nonumber \\ 
 - \mathrm{i}\, A_1\, \theta \ket*{\phalf} \otimes \left( \alpha \ket*{\mhalf} + \beta \ket*{\phalf} \right) \:.
\label{eqn-proj-decoded}
\end{eqnarray}
Thus, a projective measurement of the electron spin state $m_S$ yields whether an error occurred ($m_S=+1/2$) or not ($m_S=-1/2$); in either case, the error-corrected qubit state is recovered on a $m_I=\pm1/2$ superposition. 

There are certain requirements on a physical system on which this protocol is to be implemented. From the fundamental perspective, we require: that $S\geq1/2$ and $I\geq3/2$; that all transitions are independently spectrally addressable; and the availability of projective measurement of the electron spin. 

While a number of physical systems show promise in meeting these requirements~\cite{Sigillito2017-my, Vincent2012-mi, Omanakuttan2021-qm}, we have chosen to explore the implementation and spin dynamics of the protocol in a system offering experimental convenience and flexibility, at the cost of the availability of projective measurement. 

Manganese defects (Mn$^{2+}$, atomic configuration [Ar]3d$^5$4s$^0$) in single-crystal Wurtzite zinc oxide (ZnO) offer a homogeneous ensemble of highly coherent electronic spins with $S = 5/2$, each hyperfine-coupled to the $^{55}$Mn nuclear spin $I = 5/2$, yielding a spin Hilbert space of dimension 36. We can conveniently control the electron-nuclear spin system using electron and nuclear pulsed magnetic resonance, and detect ensemble electron spin coherences via Hahn echoes. The experiment reported here does not require additional optical excitation or readout processes~\cite{Waldherr2014-oo,taminiau2014universal, robledo2011high, bradley2019ten, ruskuc2022nuclear}.

\begin{figure*}
\includegraphics[width=\textwidth]{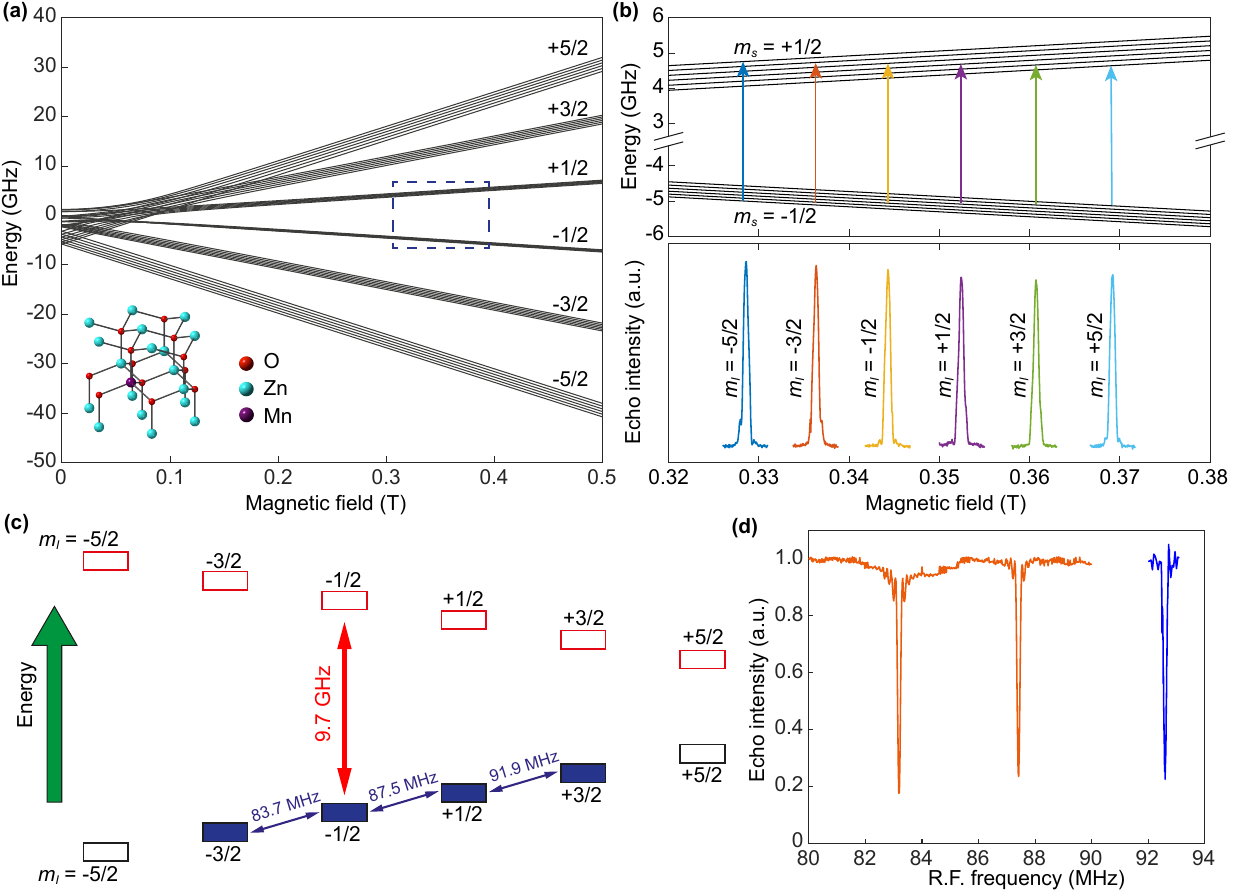}
\caption{ESR and ENDOR spectroscopy at a temperature of 30~K where the electronic $T_{1\mathrm{e}} \approx 1.3$~ms. (a) The spin energy levels of Mn$^{2+}$ defects in ZnO. (Inset) the atomic structure of the defect. (b) Echo-detected field swept ESR spectra at 9.7~GHz, showing the six $\ket{-1/2, m_I}$ to $\ket{1/2 , m_I}$ ESR transitions corresponding to each $m_I$ projection. (c) Owing to the second order hyperfine shifts, the nuclear spin transitions within $m_S=-1/2$ are non-degenerate and therefore spectrally addressable. (d) Davies ENDOR spectra obtained at 0.3443~T. The two peaks at 83.7~MHz and 87.5~MHz (orange) correspond to transitions from $\ket{-1/2, -1/2}$ to $\ket{-1/2, -3/2}$ and to $\ket{-1/2, +1/2}$ respectively. The peak at 92.6~MHz, corresponding to the transition from $\ket{-1/2, +1/2}$ to $\ket{-1/2, +3/2}$, is obtained from a modified Davies ENDOR sequence including a RF $\pi$-pulse at 87.5~MHz before the frequency-swept RF pulse, as described in Section~III of the Supplementary Information.}
\label{ESRandENDOR}
\end{figure*}

The electron spin resonance (ESR) and electron-nuclear double resonance (ENDOR) spectra of Mn:ZnO have been studied extensively~\cite{HAUSMANN19681369,george2013coherent, bottcher201455mn}. For a magnetic field $B_z$ aligned along the crystal $c$-axis, the low-energy spin degrees of freedom are described well by the Hamiltonian
\begin{equation}
H = (\gamma_S \hat{S_z} + \gamma_I \hat{I_z} ) B_z + A_{\mathrm{h.f.}} \hat{\bf S}\cdot \hat{\bf I} - D \hat{S_z}^2 
\label{H}
\end{equation}
comprising  electron and nuclear spin Zeeman terms, an isotropic hyperfine coupling, and an electron spin zero field splitting. In  units convenient for understanding the experiment, $\gamma_S/h \approx 28.02$~GHz/T, $\gamma_I/h \approx -10.96 $~MHz/T, $A_{\mathrm{h.f.}}/h \approx -220.0 $~MHz, and $D/h \approx 707.0$~MHz, yielding an energy level structure shown as a function of magnetic field in Fig.~\ref{ESRandENDOR}(a). The electron spin coherence time $T_\mathrm{2e}$ is about 80~$\mu$s below 30~K and approaches a millisecond below 10~K~\cite{george2013coherent} (see Supplementary Information Section II).  The electron spin zero field splitting ensures that the six allowed ESR transitions (for which $\Delta m_S = \pm 1$) are non-degenerate. The $A_{xx}$ and $A_{yy}$ terms in the isotropic hyperfine interaction lead to second-order shifts in the nuclear spin energy levels of order $A_{\mathrm{h.f.}}^2 / \gamma_S B_z$, which for an electron spin Zeeman frequency in X-band ($\sim$9.7~GHz, corresponding to $B_z\sim 0.35$~T) lifts the degeneracy of the allowed NMR transitions ($\Delta m_I = \pm 1$) by about 4~MHz. Thus, owing to the final two terms in the Hamiltonian, Eqn.~\ref{H}, all of the spin transitions occur at distinct frequencies and are spectrally addressable (see Supplementary Information Sections~III, IV).

In this paper, we exploit this addressability to select the subspace suitable for implementation of the fault-tolerant memory protocol described above. In fact, the encoding in Eqns.~\ref{eqn-electron-qubit}, \ref{eqn-encoded-state} and \ref{eqn-encoding} and the experiments described below involve only one level from the $m_S=+1/2$ subspace, $\ket{+1/2}\otimes\ket{-1/2}$, in addition to the four levels in the $m_S=-1/2$ subspace, $\ket{-1/2}\otimes\ket{m_I}$ with $m_I = -3/2, -1/2, +1/2, +3/2$. In our finite-temperature ensemble experiment, this approach provides an important practical advantage: the  electronic Zeeman energy is significantly larger than the nuclear state energy splittings in the $m_S = -1/2$ subspace, so our thermal state is effectively a pseudo-pure state, not requiring active preparation~\cite{knill1998effective, knill2000algorithmic} (see Supplementary Information Section~V). {A further practical reason for working within the $m_S = \pm 1/2$ subspace is that these levels are unaffected by the dominant inhomogeneous broadening mechanism in this material, which is the strain in the zero-field splitting parameter, $D$ in Eqn.~\ref{H}~\cite{george2013coherent}.}

We mounted a Mn doped ZnO single crystal, of dimensions of 2.5 mm $\times$ 6 mm $\times$ 0.5 mm, in a Bruker MD4 resonator with the crystal $c$-axis parallel to the external magnetic field. 
We measured ESR and ENDOR spectra at 30~K (elevated from the base temperature so as to limit $T_\mathrm{1n}$ and therefore increase the available experimental shot repetition rate) using an X-band ($\sim$9.7~GHz) home-built spectrometer. Through ESR (Fig.~\ref{ESRandENDOR}(b)) and ENDOR (Fig.~\ref{ESRandENDOR}(d)) spectroscopies, we obtained the frequencies of the transitions within the relevant subspace. Subsequent coherence transfer experiments between the electronic and nuclear degrees of freedom \cite{morton2008solid, brown2011coherent, wolfowicz2015coherent} and within the nuclear spin qudit subspace confirmed that quantum information can be generated, encoded, and read out in the electron-nuclear coupled system (see Supplementary Information Sections~III, IV). The fidelities of the relevant microwave and RF pulses are estimated~\cite{Morton2005} to be $99.5\pm0.1$\% and $93.5\pm0.1$\%, with details provided in Supplementary Information Section~VI.

\begin{figure*}
\includegraphics[width=0.8\textwidth]{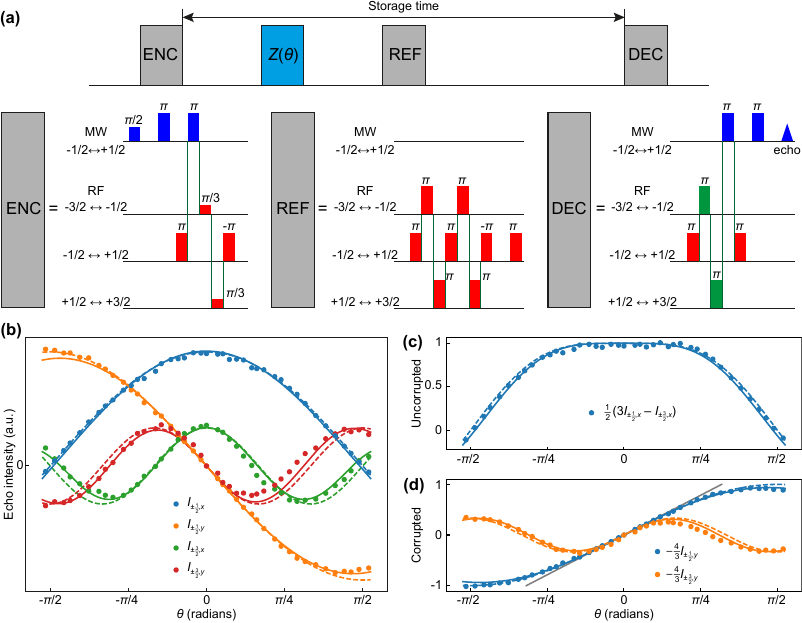}
\caption{Implementation and evolution of the error-protected state. (a) The pulse sequence comprising: ``ENC'', the generation of a qubit state and its encoding in the logical-qubit subspace; ``$\widehat{Z(\theta)}$'', the controlled artificial perturbation; ``REF'', refocussing the inhomogeneities within the nuclear subspace; ``DEC'', decoding pulses allowing measurement of the nuclear coherences $m_I=-1/2,+1/2$ (green pulses absent) and $m_I=-3/2,+3/2$ (green pulses present). {All MW pulses are resonant with the $m_S = \pm 1/2$ transition for the $m_I = -1/2$ nuclear spin projection. All RF pulses are within the $m_S=-1/2$ subspace.} (b) The evolution of the four detectable echoes at the end of the sequence as a function of the applied perturbation angle $\theta$ {with a fixed storage time of 0.1~ms}. (c) A linear combination of the echoes yielding, for small $\theta$, the squared amplitude of the uncorrupted component $A_0^2$. (d) Rescaled appropriately, the out of phase echoes yield the amplitude of the corrupted component for small $\theta$. The grey line has gradient $1$ and passes through the origin. In (b), (c) and (d): points are data; solid lines represent the global fit to a model with six parameters, $A_0, \cdots, A_5$, as defined in Eqn.~\ref{eqn-errored-state}; dashed lines represent the ideal exact evolution of the system with the only parameter being the overall vertical scale.}
\label{ErrCorr}
\end{figure*}

The coherence transfer experiments equip us to implement the encoding defined in Eqns.~\ref{eqn-encoded-state} and~\ref{eqn-encoding}, and to explore the dynamics of the encoded state. Fig.~\ref{ErrCorr}(a) shows the pulse sequence on which our investigation is based; it comprises four units. The first, the encoding unit labelled ``ENC'', generates an electron spin qubit state (chosen as $\alpha=\beta=1/\sqrt{2}$, i.e.\ a $\hat{\sigma_x}$ eigenstate) and then transfers it to the logical states defined in Eqns.~\ref{eqn-encoded-state} and~\ref{eqn-encoding}. 

This is followed by a controlled application of the perturbation against which the encoding is designed to protect, a fluctuation of $B_z$, in a unit of the sequence labelled ``$\widehat{Z(\theta)}$''. We achieve this by applying a current pulse of calibrated amplitude and duration to a pair of appropriately oriented Helmholz coils proximal to the sample (see Supplementary Information Section~VII). With small currents applied for short durations, the effect of the artificial perturbation is small and the approximation in Eqn.~\ref{eqn-errored-state} holds. By increasing the current and the pulse duration, we have the scope to explore experimentally the regime beyond the small-perturbation limit; our apparatus allows us to apply pulses for which $\theta$ exceeds $\pm\pi/2$~radians.

Working with an ensemble introduces a requirement that we account for inhomogeneities. In the qubit case, where only two energy eigenstates are involved in a superposition, the effects of static inhomogeneities are refocused by a single $\pi$-pulse. In our qudit case, involving a superposition across four energy eigenstates, the situation is more subtle; it is not straightforward to implement an operation achieving full refocussing while preserving the system within the error-protected subspace. (Approaches to refocusing arbitrary qudit states have been identified \cite{vitanov2015dynamical}, but they lead to extended periods when the encoded state resides outside the logical qubit space, destroying the protoection.) Under this constraint it is only possible to guarantee refocussing within the $m_I = \pm3/2$ and $m_I = \pm1/2$ subspaces, but the relative phase between these subspaces is uncontrolled. This refocussing is achieved by the third unit in the sequence, comprising eight $\pi$-pulses and labelled ``REF'' (see Supplementary Information Section~VIII).

The experiment therefore yields, as a function of the artificially applied error angle $\theta$, four echoes: the in phase ($x$) and out of phase ($y$) components of echoes arising from the $m_I=+1/2,-1/2$ (obtained with the green pulses in Fig.~\ref{ErrCorr}(a) absent) and the $m_I=+3/2,-3/2$ coherences (obtained with the green pulses present). We label the echo amplitudes $I_{\pm\frac{1}{2},x}$, $I_{\pm\frac{1}{2},y}$, $I_{\pm\frac{3}{2},x}$, $I_{\pm\frac{3}{2},y}$.
These are plotted in Fig.~\ref{ErrCorr}(b) as points for a fixed storage time of 0.1~ms. 
They are accompanied by a global fit to a model based on a truncation of the series in Eqn.~\ref{eqn-errored-state} with free parameters $A_n$, $n = 0 \cdots 5$ (solid lines). Also shown are the amplitudes expected under the exact theoretical evolution with $\theta$ (dashed lines). The values of the fit parameters are given in Table~\ref{table-As} normalised to $A_0$, accounting for the overall echo signal gain in the experiment. The proximity of $A_n/A_0$ (with $n>0$) to~$1$ is indicative of how close the evolution of the system is to the exact theoretical evolution.

\begin{table}
\caption{Values of the fit parameters $A_n$. They are normalised to $A_0$, which defines the overall vertical scale in Fig.~\ref{ErrCorr}(b). Quoted errors are $\pm1\sigma$ bounds.}
\begin{tabular}{|c|c|c|c|c|c|}
\hline
$A_0$ & $A_1/A_0$ & $A_2/A_0$ & $A_3/A_0$ & $A_4/A_0$ & $A_5/A_0$ \\
\hline
$14.19$ & $0.997$ & $1.12$ & $1.13$ & $1.02$ & $1.2$ \\
$\pm0.05$ & $\pm0.008$ & $\pm0.02$ & $\pm0.05$ & $\pm0.05$ & $\pm0.2$ \\
\hline
\end{tabular}
\label{table-As}
\end{table}

Given that we cannot achieve full refocussing and that our detection of the properties of the system state is via Hahn echoes of electron spin coherences, the question arises: what features of the protected state can we observe as a function of the perturbation $\widehat{Z(\theta)}$? 
The echo amplitudes can be expressed in terms of $A_n$ as
\begin{eqnarray}
I_{\pm\frac{1}{2},x} &=& \frac{3}{4} A_0^2 - \frac{3\theta^2}{16}(A_1^2+A_0 A_2) + O(\theta^4) \nonumber \\
I_{\pm\frac{1}{2},y} &=& -\frac{3\theta}{4} A_0 A_1  + \frac{\theta^3}{32} (3 A_1 A_2 + A_0 A_3)  +O(\theta^5) \nonumber \\
I_{\pm\frac{3}{2},x} &=&  \frac{1}{4}A_0^2 - \frac{9\theta^2}{16}(A_1^2+A_0 A_2)  +O(\theta^4) \nonumber \\
I_{\pm\frac{3}{2},y} &=& -\frac{3\theta}{4} A_0 A_1  + \frac{9\theta^3}{32} (3 A_1 A_2 + A_0 A_3)  +O(\theta^5). \nonumber \\
\end{eqnarray}
(We note here that if we had chosen our original qubit to be an eigenstate of $\hat{\sigma_y}$, the $x$ and $y$ components of these echoes would be exchanged.)

Our principal experimental objective is to monitor the evolution of state amplitude from the logical qubit subspace into the orthogonal ``error'' subspace defined immediately following Eqn.~\ref{eqn-errored-state} under the action of the perturbation, i.e.\ to monitor the amplitudes $A_0$ and $A_1\,\theta$ as a function of $\theta$. We can access these quantities from the available echo data. The linear combination
\begin{equation}
\frac{3 I_{\pm\frac{1}{2},x} - I_{\pm\frac{3}{2},x}}{2} = A_0^2 + O(\theta^4)
\label{echo-lin-comb-Asquare}
\end{equation}
yields the square of the uncorrupted amplitude, and is plotted in Fig.~\ref{ErrCorr}(c) as a function of $\theta$ (points), along with the model fit (solid line) and the predicted exact evolution (dashed line). The remarkably flat behaviour over a wide range around $\theta=0$ is a direct reflection of the fact that the perturbation is moving amplitude from the logical qubit subspace into an orthogonal subspace. We may obtain the amplitude of the corrupted state directly from a scaling of the out of phase echos
\begin{eqnarray}
-\frac{4 I_{\pm\frac{1}{2},y}}{3} = A_0 A_1 \theta + O(\theta^3)   \nonumber \\
 -\frac{4 I_{\pm\frac{3}{2},y}}{3} = A_0 A_1 \theta + O(\theta^3)
\label{echo-scale-error-ampl}
\end{eqnarray}
as plotted in Fig.~\ref{ErrCorr}(d). Over a significant range around $\theta = 0$, both quantities are coincident with a line of gradient 1 through the origin, confirming that they faithfully report the corrupted amplitude. 

Given the coincidence of $T_{2\mathrm{n}}$ and $T_{1\mathrm{e}}$ obtained from coherence transfer experiments (See Supplementary Section III), we expect the dominant natural decoherence process affecting the nuclear spin to be fluctuations of the magnetic field along the quantisation axis arising from electron spin flips coupled through the diagonal component of the hyperfine interaction. It is therefore interesting to explore the behaviour of the encoded state under the action of the natural decoherence processes present in the system, i.e., as a function of the storage time without the application of the artificial perturbation. 

In contrast to our artificial perturbation operation $\widehat Z(\theta)$, natural processes affect each member of the ensemble independently and incoherently. This means that for our choice of qubit state all out-of-phase echo components are zero and we cannot monitor the corrupted amplitude using the strategy in Eqn.~\ref{echo-scale-error-ampl}. Instead, we construct a linear combination from the in phase echoes
\begin{equation}
\frac{2 \left( I_{\pm\frac{1}{2},x} - 3 I_{\pm\frac{3}{2},x} \right) }{3} = (A_1^2 + A_0 A_2) \theta^2
\label{echo-lin-comb-err-square}
\end{equation}
which gives us a quantity related to the square of the amplitude of the corrupted component.

\begin{figure}
\includegraphics[width=0.5\columnwidth]{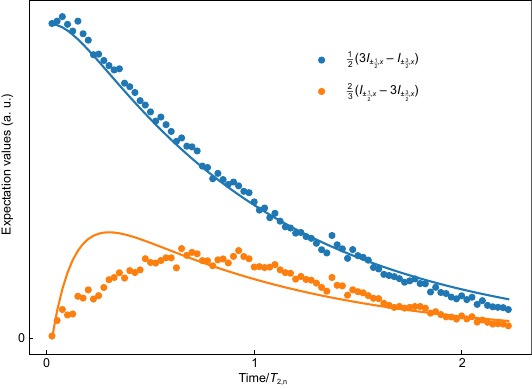}
\caption{Evolution of the encoded state with natural decoherence. Points are the amplitude squared of the uncorrupted component (Eqn.~\ref{echo-lin-comb-Asquare}, blue) and the quantity related to the squared amplitude of the corrupted component (Eqn.~\ref{echo-lin-comb-err-square}, orange). Lines correspond to a minimal master equation model incorporating only $\hat{I_z}$ as a relaxation operator.}
\label{natural-decoh}
\end{figure}

In Fig.~\ref{natural-decoh} we plot this quantity and the square of the uncorrupted amplitude (Eqn.~\ref{echo-lin-comb-Asquare}) as a function of the storage time (defined in Fig.~\ref{ErrCorr}(a)) without applying the artificial perturbation $\widehat{Z(\theta)}$. At short times (below $\sim 0.1 \, T_{2\mathrm {n}}$) the uncorrupted component is approximately independent of time, after which it falls approximately exponentially with a timescale characteristic of $T_{2\mathrm {n}}$. The corrupted component starts at zero and increases linearly at short times, eventually decaying at longer times also on the $T_{2\mathrm {n}}$ timescale. 
This behaviour is consistent with our logical qubit state being orthogonal to the state to which the dominant phase relaxation process mixes. The solid lines in Fig.~\ref{natural-decoh} represent the predictions of a minimal Lindblad master equation model with the relaxation operator $\sqrt{2/T_{2\mathrm {n}}}\hat{I_z}$ determined from an independent measurement of $T_{2\mathrm {n}}$; the only free parameter is the vertical scaling. This model describes the qualitative behaviour well.

The results presented in Figs.~\ref{ErrCorr}, ~\ref{natural-decoh} offer strong evidence that we can implement the logical qubit encoding presented in Eqn.~\ref{eqn-encoding}, and that its dynamics under the perturbation that it is designed to protect against are as expected. We conclude that the deployment of hyperfine-coupled nuclear qudits can indeed offer a valuable resource for implementing fault-tolerant quantum memories~\cite{chiesa2020molecular,Gross2021-qx,Lim2023-yy}. The potential of this approach provides fresh impetus for developing single-spin, single-shot projective measurement for condensed matter based electron spin qubits, including broadening the scope of applicability of demonstrated methods~\cite{Thiele2014-il,Vincent2012-mi} as well as exploring new approaches. The important question of the potential performance of a quantum memory of the kind described here is explored elsewhere~\cite{chiesa2020molecular}, but we note that it will depend critically on the fidelities of both the encoding and measurement protocols. 

%

\section*{Acknowledgments}
This project has received funding from the European Union's Horizon 2020 research and innovation programme under grant agreements 862893 (FATMOLS) and 863098 (SPRING). JL acknowledges support from the Royal Society under grant URF$\backslash$R1$\backslash$201132. MV is grateful to the Hill Foundation for financial support.

\newpage

\end{document}



\title{Supplementary Information: Demonstrating experimentally the encoding and dynamics of an error-correctable logical qubit on a hyperfine-coupled nuclear spin qudit}


\author{Sumin Lim}
\affiliation{CAESR, Department of Physics, University of Oxford, The Clarendon Laboratory, Parks Road, Oxford OX1 3PU, UK}
\author{Mikhail Vaganov}
\affiliation{CAESR, Department of Physics, University of Oxford, The Clarendon Laboratory, Parks Road, Oxford OX1 3PU, UK}
\author{Junjie Liu}

\affiliation{CAESR, Department of Physics, University of Oxford, The Clarendon Laboratory, Parks Road, Oxford OX1 3PU, UK}
\author{Arzhang Ardavan}
\affiliation{CAESR, Department of Physics, University of Oxford, The Clarendon Laboratory, Parks Road, Oxford OX1 3PU, UK}






\maketitle

\section{Operations encoding and decoding the logical qubit}
\label{nuclear-enc-dec}

In this section we present the detailed operations by which the qubit state in Eqn.~2 of the main manuscript is rotated onto the the logical qubit state in Eqn.~3, and the decoding operation rotating the state in Eqn.~6 to the state in Eqn.~7. In quantum spin systems, controlled rotations between states with $\Delta m_S = \pm 1$ or $\Delta m_I = \pm 1$ may be achieved using magnetic resonance pulses; in specific circumstances~\cite{GeorgePRL2013, Asaad2020-fx} $\Delta m_S = \pm 2$ or $\Delta m_I = \pm 2$ transitions are also available. The protocol described here depends on each of these transitions being spectrally addressable, as is the case, for example, in the Mn:ZnO system.

In a two-state system, the basic operation is a rotation 
\begin{equation}
R_{\ket*{\mhalf} \leftrightarrow \ket*{\phalf},\theta} = \exp\left(-\frac{\mathrm{i} \theta \sigma_y}{2}\right) = 
\left(
\begin{array}{cc}
\cos\left(\frac{\theta}{2}\right) & -\sin\left(\frac{\theta}{2}\right) \\
\sin\left(\frac{\theta}{2}\right) & \cos\left(\frac{\theta}{2}\right)
\end{array}
\right)
\end{equation}
where the axis of the rotation (which, here, is the $y$-axis in the rotating frame) is chosen by adjusting the phase of the magnetic resonance pulse, and the angle of the rotation, $\theta$, is controlled by adjusting the duration or the amplitude of the pulse.

\subsection{Encoding and decoding within the $m_S = -\frac{1}{2}$ subspace}
\label{nuclear-enc-dec}

At the heart of the protocol is a unitary transformation within the $m_S = -\frac{1}{2}$ subspace,
\begin{eqnarray}
U_\mathrm{enc} & = & \ket*{0_L}  \bra*{\mhalf,\mhalf}  \nonumber \\
& + & \ket*{1_L}  \bra*{\mhalf,\phalf}  \nonumber \\
& + & \ket*{I_z 0_L}  \bra*{\mhalf,\mthalf} \nonumber \\
& + &  \ket*{I_z 1_L}  \bra*{\mhalf,\pthalf} 
\end{eqnarray}
where $\ket{m_S, m_I} \equiv \ket{m_S} \otimes \ket{m_I}$, $\ket{0_L}$ and $\ket{1_L}$ are defined in Eqn.~4 of the main manuscript, and 
\begin{eqnarray}
\ket*{I_z 0_L} & = & \frac{2}{\sqrt{3}} \hat{I}_z \ket*{0_L} \nonumber \\
\ket*{I_z 1_L} & = & \frac{2}{\sqrt{3}} \hat{I}_z \ket*{1_L} 
\end{eqnarray}
with factors of $2/\sqrt{3}$ included for normalisation. $U_\mathrm{enc}$ encodes a nuclear spin qubit
\begin{equation}
\alpha \ket*{\mhalf,\mhalf} + \beta \ket*{\mhalf,\phalf}
\label{SI_nuc_qubit}
\end{equation}
to the state $\ket{\psi_L}$ given in Eqn.~3 of the main manuscript.

Under the action of the perturbation, $\ket{0_L}$ and $\ket{1_L}$ are mixed with $\ket*{I_z 0_L}$ and $\ket*{I_z 1_L}$ respectively. 

The inverse transformation, $U_\mathrm{enc}^\dagger$, decodes from the protected basis so that the uncorrupted component is restored to the state in Eqn.~\ref{SI_nuc_qubit}, and the error component arrives on the superposition
\begin{equation}
\epsilon \left( \alpha \ket*{\mhalf,\mthalf} + \beta \ket*{\mhalf,\pthalf} \right)
\label{SI_nuc_qubit_err}
\end{equation}
where, comparing with Eqns.~6 and 7 of the main manuscript, $\epsilon = -\mathrm{i}\theta$.

The practical implementation of $U_\mathrm{enc}$ depends on the operations that are available in the specific physical system on which the protocol is to be implemented. For a system in which $\Delta m_I = \pm 1, \pm 2$ transitions are available, an efficient operation yielding $U_\mathrm{enc}$ is provided by the sequence of pulses represented by operators (to be read from right to left)
\begin{equation}
R_{\ket*{\mhalf}  \leftrightarrow {\ket*{\pthalf}}, \frac{\pi}{3} }
\cdot R_{\ket*{\mthalf } \leftrightarrow \ket*{\phalf}, \frac{5 \pi}{3}}
\cdot R_{\ket*{\mhalf} \leftrightarrow \ket*{\phalf},-\pi} \:.
\end{equation}
The decoding pulse sequence implementing $U_\mathrm{enc}^\dagger$ is generated by reversing the sequence of pulses and replacing $\theta$ with $-\theta$ for each. 

In physical systems for which only $\Delta m_I = \pm 1$ transitions are available, alternative equivalently functional encodings within this space can be identified.

\subsection{Swapping between electron and nuclear spin states}

The encoding between main manuscript Eqns.~2 and 3 requires a step before the operation described in Section~\ref{nuclear-enc-dec}, transferring the electron qubit state in Eqn.~2 of the main manuscript onto the nuclear qubit Eqn.~\ref{SI_nuc_qubit}. This operation can be achieved using a sequence of one nuclear $\pi$ pulse and one electron $\pi$ pulse~\cite{Morton2008-gt},
\begin{equation}
\label{e_to_n}
R_{\ket*{\mhalf,\phalf} \leftrightarrow \ket{\phalf,\phalf}, \pi} 
\cdot R_{\ket*{\phalf,\mhalf} \leftrightarrow \ket{\phalf,\phalf}, -\pi} \: .
\end{equation}

Following the decoding transformation described in Section~\ref{nuclear-enc-dec}, the uncorrupted component is already in the superposition given by the first term in main manuscript Eqn.~7. The final decoding step is to transfer the corrupted component from the 
$\ket*{\mhalf,\mthalf},\ket*{\mhalf,\pthalf}$ superposition to the 
$\ket*{\phalf,\mhalf},\ket*{\phalf,\phalf}$ superposition. This is achieved through a pair of electron $\pi$ pulses followed by a pair of nuclear $\pi$ pulses,
\begin{equation}
R_{\ket{\phalf,\phalf} \leftrightarrow \ket{\phalf,\pthalf},\pi}
\cdot R_{\ket*{\phalf,\mthalf} \leftrightarrow \ket{\phalf,\mhalf}, -\pi}
\cdot R_{\ket*{\mhalf,\pthalf} \leftrightarrow \ket{\phalf,\pthalf},-\pi}
\cdot R_{\ket*{\mhalf,\mthalf} \leftrightarrow \ket{\phalf,\mthalf},-\pi} \:.
\end{equation}

\subsection{Our experimental encoding stage}

Owing to the lack of projective measurement in our apparatus, there is no value in our experiment implementing the full decoding described here. Instead we use a simpler but equivalent encoding, and we explore the dynamics of the encoded state by detecting various coherences as described in the main manuscript and in Section~\ref{State}.

\section{Low-temperature relaxation for the electron spins}

\begin{figure}[htb!]
\includegraphics[width=12cm]{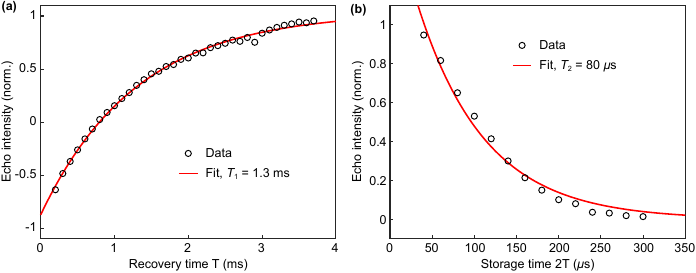}
\caption{The measurement of \textbf{(a)} the spin-lattice relaxation time $T_{1\mathrm{e}}$ and \textbf{(b)} the phase coherence time $T_{2\mathrm{e}}$.}
\label{fig_electron_relaxation}
\end{figure}

The spin-lattice relaxation time ($T_{1\mathrm{e}}$) for the electron is measured employing the standard inversion recovery sequence  [$\pi - T - \pi/2 - \tau - \pi - \mathrm{echo}$], recording the echo intensity vs. $T$. The phase coherence time ($T_{2\mathrm{e}}$) for the electron spins is measured by the standard Hahn-echo sequence [$\pi/2 - T - \pi - T - \mathrm{echo}$] and fitting the echo intensity vs. the coherence storage time $2T$. The results are shown in Fig.~\ref{fig_electron_relaxation}. The measurements are performed at 30~K with the static magnetic field of 0.3433~T, which corresponds to the resonance condition for the ESR transition between the $m_S = \pm 1/2$ states with $m_I = -1/2$. Both data are well described by single exponential functions, yielding $T_{1\mathrm{e}} = 1.3 \pm 0.045$ ms and $T_{2\mathrm{e}} = 80 \pm 4.05 \,\mu$s.

\section{Electron-nuclear double resonance (ENDOR) and coherence transfers of quantum information between electron and nuclear spins}

\begin{figure}[htb!]
\includegraphics[width=\columnwidth]{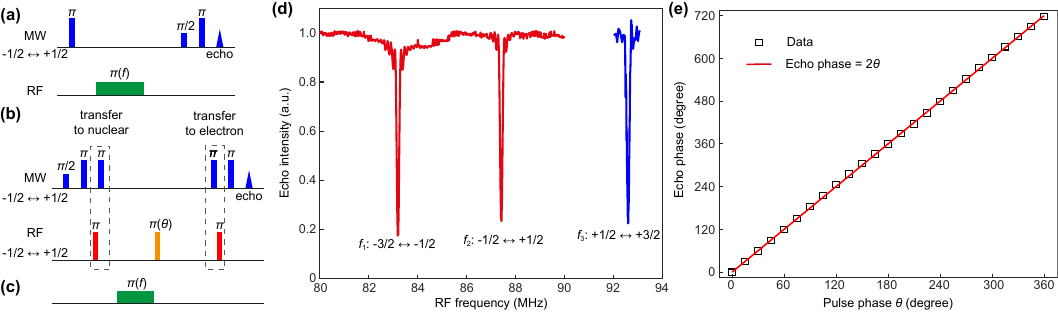}
\caption{\textbf{(a)} the standard Davies ENDOR sequence used to measure the allowed ($\delta m_I = \pm1$) nuclear spin resonance from the $\ket{-1/2}\otimes\ket{-1/2}$ state. The echo intensity is recorded while varying the frequency of the RF pulse $\pi{(f)}$ to match the resonance conditions. \textbf{(b)} the pulse sequence that implements the transfer of quantum coherence between the electron and nuclear spins. \textbf{(c)} an additional RF pulse $\pi(f)$ is inserted to the coherence transfer sequence in \textbf{(b)}, allowing the detection of the $\ket{-1/2}\otimes\ket{+1/2} \leftrightarrow \ket{-1/2}\otimes\ket{+3/2}$ resonance frequency by tuning the frequency of $\pi(f)$. \textbf{(d)} the echo intensity as a function of the frequency of the pulse $\pi(f)$. The red curve is measured using the sequence in \textbf{(a)} and the blue trace is recored using the sequence described in \textbf{(b)} and \textbf{(c)}. \textbf{(e)} the phase of the final electron spin echo as a function of the phase of the refocusing pulse $\pi(\theta)$ in \textbf{(b)}, confirming the measured coherence is stored in the nuclear spins.}
\label{fig_endor}
\end{figure}

Direct measurement of the Mn nuclear spins is only possible at very high doping levels which significantly compromises the coherence time of the electron spins. Therefore, we apply electron-nuclear double resonance techniques to control both the electron and nuclear spins, and measure the nuclear spins by means of electron spin echoes. The protocol for encoding the logical qubit utilises the nuclear spin subspace $m_I = \pm1/2$ and $\pm3/2$ (all with $m_S = -1/2$). Hence, it is essential that we are able to generate superpositions between any nuclear spin states in this subspace using RF control pulses. This requires access to three nuclear spin transitions, $f_1 : \ket{-1/2}\otimes\ket{-1/2} \leftrightarrow \ket{-1/2}\otimes\ket{+1/2}$, $f_2 : \ket{-1/2}\otimes\ket{-1/2} \leftrightarrow \ket{-1/2}\otimes\ket{-3/2}$ and $f_3 : \ket{-1/2}\otimes\ket{+1/2} \leftrightarrow \ket{-1/2}\otimes\ket{+3/2}$. (All states are represented using the $\ket{m_S}\otimes\ket{m_I}$ notation). In the experiment, the static magnetic field $B_0$ is fixed at 0.3443~T, which corresponds to the resonance field of the $\ket{-1/2}\otimes\ket{-1/2} \leftrightarrow \ket{+1/2}\otimes\ket{-1/2}$ ESR transition. This combination of  MW frequency and  magnetic field allows us to access $f_1$ and $f_2$ nuclear spin transitions directly using the standard Davies ENDOR sequence [Fig.~\ref{fig_endor}(a)]. The result is shown in the red curve in Fig.~\ref{fig_endor}(d). When sweeping the frequency of the RF  $\pi{(f)}$-pulse, two resonance peaks are observed, corresponding to the two allowed nuclear spin transitions involving the $\ket{-1/2}\otimes\ket{-1/2}$ state.

Accessing the $f_3 : \ket{-1/2}\otimes\ket{+1/2} \leftrightarrow \ket{-1/2}\otimes\ket{+3/2}$ transition is more challenging. In order to measure the nuclear transition frequency $f_3$, we first introduce the coherence transfer scheme used in this work. This scheme, together with its variations, allow us to transfer electron coherence to the nuclear spins and detect it in a highly diluted sample. Fig.~\ref{fig_endor}(b) shows the pulse sequence used to transfer, store and measure the coherence between $m_I = \pm 1/2$ states. A coherence is first generated by the initial MW $\pi/2$  pulse, placing the electron spins into the superposition of $\ket{-1/2}\otimes\ket{-1/2}$ and  $\ket{+1/2}\otimes\ket{-1/2}$ states. The following MW $\pi$ pulse refocusses the electron spins to mitigate the free induction decay due to inhomogeneities, allowing the coherence to be transferred to the nuclear spins by the following RF and MW $\pi$ pulses indicated by the dashed box. The coherence generated by the first MW $\pi/2$ pulse is now completely stored in the nuclear spin, between the superposition of the $\ket{-1/2}\otimes\ket{-1/2}$ and $\ket{-1/2}\otimes\ket{+1/2}$ states. This quantum coherence may be stored for an extended period (of the time scale of the nuclear $T_{2\mathrm{n}}$). A RF refocussing pulse, $\pi(\theta)$ of phase of $\theta$ is applied at the halfway point of the storage time to remove the effects of inhomogeneities associated with the nuclear spin, allowing the remaining coherence to be transferred back to the electron spins and measured using the electron spin echo.

A four-step phase cycling of the initial MW $\pi/2$ and the RF $\pi$ refocussing pulses (Table.~\ref{phase_cycle}) is implemented to remove  undesired echo contributions, including the stimulated electron spin echo (which decays on the timescale of $T_{1e}$), electron echo/free induction decay due to the last two MW $\pi$ pulses, and the detector DC offset, ensuring the measured echo signal is purely due to the coherence transfer via the nuclear spins. This is further verified by varying the phase of the RF refocusing pulse $\pi(\theta)$. Only the coherence stored in the nuclear spins is expected to gain an additional phase factor of $2\theta$, which can be measured in the final electron spin echo, whereas all other undesired echo contributions are insensitive to $\theta$. The data in Fig.~\ref{fig_endor}(e) match with the $2\theta$ expectation perfectly, confirming all the detect signals are transferred via the electron-nuclear-electron pathway.
\begin{table}[htb!]
\centering
\caption{The phase cycling sequence and the signal for the coherence transfer scheme.}
\label{phase_cycle}
    \begin{tabular}{ | c | c | c | c | } 
    \hline
     \enspace Step number \quad&\enspace MW $\pi/2$ pulse \quad&\enspace RF refocusing pulse \quad&\enspace Signal \quad\\
     \hline
    1 & $+x$ & $+x$ & $+$ \\
    2 & $-x$ & $+x$ & $-$ \\
    3 & $+x$ & $+y$ & $-$ \\
    4 & $-x$ & $+y$ & $+$ \\
    \hline
    \end{tabular}%
\end{table}%

Finally, we present the measurement of the $\ket{-1/2}\otimes\ket{+1/2} \leftrightarrow \ket{-1/2}\otimes\ket{+3/2}$ transition frequency using a modified ENDOR sequence. The sequence is largely identical to the coherence transfer scheme [Fig.~\ref{fig_endor}(b)] with an additional RF pulse $\pi(f)$ applied between the first two RF $\ket{-1/2}\otimes\ket{-1/2} \leftrightarrow \ket{-1/2}\otimes\ket{+1/2}$ pulses [Fig.~\ref{fig_endor}(c)]. When the frequency of the $\pi(f)$ pulse matches that for the $\ket{-1/2}\otimes\ket{+1/2} \leftrightarrow \ket{-1/2}\otimes\ket{+3/2}$ nuclear transition, the nuclear spin coherence is transferred and stored between the superposition of the  $\ket{-1/2}\otimes\ket{-1/2}$ and $\ket{-1/2}\otimes\ket{+3/2}$ states instead. Under this circumstance, the coherence cannot be transferred back to the electron spins by the same pulse sequence, leading to a reduction of the electron spin echo, as shown by the blue curve in Fig.~\ref{fig_endor}(d). 

In summary, we successfully identified the three RF transitions: $f_ 1$ = 83.2~MHz for $\ket{-1/2}\otimes\ket{-3/2} \leftrightarrow \ket{-1/2}\otimes\ket{-1/2}$, $f_2$ = 87.4~MHz for $\ket{-1/2}\otimes\ket{-1/2} \leftrightarrow \ket{-1/2}\otimes\ket{+1/2}$ and $f_3$ = 92.6~MHz for $\ket{-1/2}\otimes\ket{+1/2}\leftrightarrow \ket{-1/2}\otimes\ket{+3/2}$. Importantly, all three transitions are non-degenerate owing to the second order hyperfine coupling shifts, allowing selective access to any desired nuclear spin states for the implementation of the fault-tolerant encoding algorithm.

\section{Time-dependence of nuclear coherences}

\begin{figure}
\includegraphics[width=\textwidth]{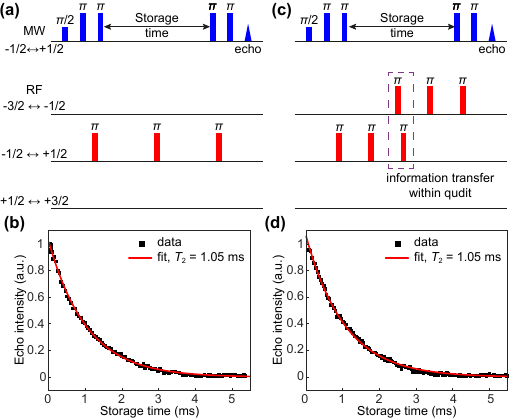}
\caption{Pulse sequences implementing coherence transfer between the electron qubit and coherences within the nuclear qudit. (a) An Electron-qubit is transferred to the nuclear $m_I = \pm 1/2$ coherence, stored for a time, and then transferred back to the electron qubit where it is read out through an electron spin echo. (b) By varying the storage time, we obtain $T_\mathrm{2n}$ from the echo decay. (c) Quantum information can be transferred between states within the qudit, in this case between the $m_I = \pm 1/2$ coherence and the $m_I = -3/2, m_I = -1/2$ coherence. By varying the storage time, we confirm (d) the same phase coherence time for each qudit transition.}
\label{CohTrans}
\end{figure}

As described in Section II, the sequence in Fig.~\ref{CohTrans}(a) generates an electronic coherence on the $m_S=-1/2,+1/2$ transition within the $m_I=-1/2$ subspace, and transfers it to a nuclear coherence on the $m_I = -1/2,+1/2$ transition within the $m_S=-1/2$ subspace, and finally restores it to the electronic state for read out via an ESR echo. In here, we measure the time dependence of the nuclear coherence decay by varying nuclear-storage-time. 

Fig.~\ref{CohTrans}(b) shows the recovered coherence amplitude as a function of the delay time, while the microwave pulse durations, the RF pulse durations, and microwave pulse intervals were kept fixed at 0.1~$ \mu$s, 7~$\mu$s, and 8~$\mu$s, respectively. The exponential decay constant yields the phase coherence time $T_{2\mathrm{n}}= 1.050 \pm 0.004$~ms for the $m_I = -1/2,+1/2$ transition. This is close to the $T_{1\mathrm{e}}$ time for the $m_S=-1/2,+1/2$ electronic transition at this temperature, indicating that phase relaxation is dominated by electronic spin flips. It exceeds very significantly the electronic $T_{2\mathrm{e}}$ time of about 80~$\mu$s, demonstrating successful coherence transfer. The phase of the detected electronic echo is controlled by the phase of the nuclear coherence refocussing pulse with the expected dependence (Fig.~\ref{fig_endor}(e)), further confirming the storage of the coherence in the nuclear states.

The next step is to demonstrate coherence transfer within the nuclear qudit. In Fig.~\ref{CohTrans}(c) we present a pulse sequence designed to transfer an electron coherence to a nuclear coherence ($m_I=-1/2,+1/2$), to transfer this coherence within the qudit (to $m_I=-1/2,-3/2$), and finally to transfer back to the electronic qubit for Hahn echo readout. By analogy with the experiment in Fig.~\ref{CohTrans}(b), measuring the dependence of the read-out echo on the storage time allows us to compare the phase coherence times for different nuclear qudit coherences. As shown in Fig.~\ref{CohTrans}(d), we find that the $m_I=-1/2,-3/2$ coherence time ($1.05 \pm 0.007$ms) matches what we found for $m_I=-1/2,+1/2$; this is exactly as we expect given that electronic $T_{1\mathrm{e}}$ processes are likely to be the dominant cause of the nuclear qudit dephasing at this temperature. 

These coherence transfer experiments allow us to calibrate the microwave and RF pulses corresponding to the operations required to implement the encoded state (Eqns.~3 and~4 in main text), with details described in Section~IV.

\section{The thermal state and state evolution}
\label{State}

In this Section, we discuss the initial thermal population and identify it as effectively a pseudo-pure state. We then explain step by step the state evolution through the experimental pulse sequence. 

\begin{figure}[htb!]
\includegraphics[width=1.0\columnwidth]{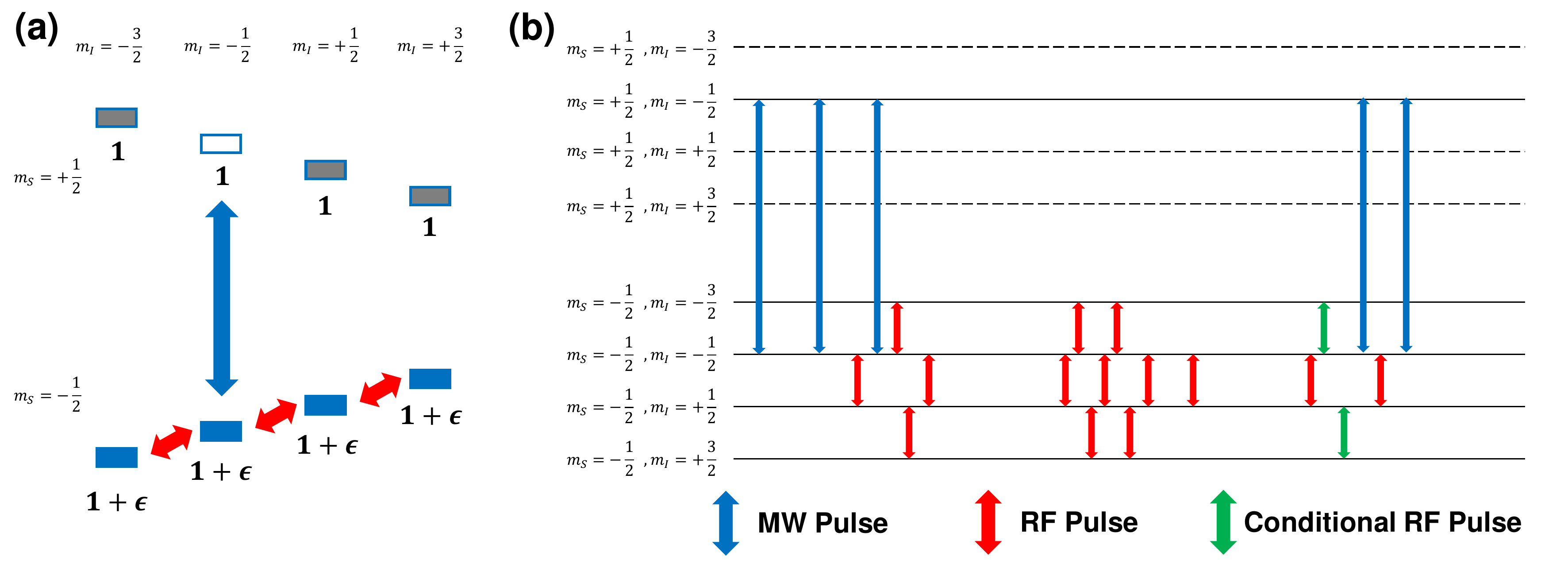}
\caption{ (a) Initial thermal population of related spin states. (b) The detail (conditional) pulse sequence for each levels. }
\label{population}
\end{figure}

\subsection{The thermal state }
\label{pseudopure-state}
In Fig.~\ref{population} (a), we indicate the initial thermal population for the relevant spin energy levels. The ground state of the system, $m_S=-5/2, m_I=-5/2$ (not shown), which will be exclusively populated for temperatures such that $k_B T$ is much smaller than any of the energy scales in the Hamiltonian, main manuscript Eqn.~8. However, for a range of practical reasons we conduct the experiment at an elevated temperature, so we must consider the thermal equilibrium population distribution in our ensemble, which represents the starting state for each experimental shot.

The population of a state of energy $E_i$ is proportional to the Boltzmann factor
$e^{-\Delta E_i / k_B T}$. In our experiment, $k_B T \sim 620$~GHz , $E_\mathrm{Zeeman} \sim 9.7$  GHz, and the nuclear transition frequencies $ \sim 0.08$ GHz. This means that the initial population difference between the $m_S = \pm 1/2$ is orders of magnitude larger than population differences within each $m_S$ manifold.
Thus to a good approximation, the density matrix of the system is
\begin{equation}
\label{population_density_matrix}
{\rho}_{\mathrm{Thermal}} = 
\begin{pmatrix}
\frac{1}{5}+\frac{\delta}{4} & 0 & 0 & 0 & 0\\ 
0 & \frac{1}{5}+\frac{\delta}{4} & 0 & 0 &0 \\ 
0 & 0 & \frac{1}{5}+\frac{\delta}{4} & 0 &0 \\ 
0 & 0 & 0 & \frac{1}{5}+\frac{\delta}{4} & 0\\ 
0 & 0 & 0 & 0 & \frac{1}{5}-\delta\\ 
\end{pmatrix}
\end{equation}
when written in the basis of states participating in the experiment,
\begin{equation}
\left\{
 \ket*{- \frac{1}{2}, -\frac{3}{2}}, \ket*{- \frac{1}{2}, -\frac{1}{2}}, \ket*{- \frac{1}{2}, +\frac{1}{2}},  \ket*{- \frac{1}{2}, +\frac{3}{2}}, \ket*{+ \frac{1}{2}, -\frac{1}{2}}
\right\} 
\label{experiment-basis}
\end{equation}
with $\ket{m_S, m_I} \equiv \ket{m_S} \otimes \ket{m_I}$.

This can be decomposed into an identity component and a component proportional to a pure state,
\begin{equation}
\label{population_density_matrix}
{\rho}_{\mathrm{Thermal}} = \left( \frac{1}{5}+\frac{\delta}{4} \right) {\bf I} 
-\frac{5 \delta}{4}
\begin{pmatrix}
0 & 0 & 0 & 0 & 0\\ 
0 & 0 & 0 & 0 &0 \\ 
0 & 0 & 0 & 0 &0 \\ 
0 & 0 & 0 & 0 & 0\\ 
0 & 0 & 0 & 0 & 1\\ 
\end{pmatrix}
\end{equation}
where ${\bf I}$ is the identity matrix and and $5\delta/4 \approx0.015$. This kind of state is known as ``pseudo-pure''~\cite{knill1998effective} because the identity component does nothing under the unitary transformations in the experiment, and the remainder is ``pure'' with a reduced amplitude of $-5\delta/4$.

\subsection{Encoding pulse sequence }

In Fig.\ref{population}(b), we illustrate the experimental pulse sequence. This is the same as the sequence in Fig~2(a) of the main manuscript, but indicating explicitly between which levels each pulse is applied. In what follows, we track the evolution of the state as the sequence progresses, represented as a vector in the basis indicated in Eqn.~\ref{experiment-basis}.

We start with the pseudo-pure state presented in Section~\ref{pseudopure-state},\begin{equation}
\begin{pmatrix}
0, & 0, &0, &0, &  1   \\
\end{pmatrix} \: .
\end{equation}
All pulses are applied with a phase generating a rotation around the $y$-axis in the rotating frame. 
The first MW $\pi / 2$ pulse generates a coherence
\begin{equation}
\begin{pmatrix}
0, & -\alpha, &0, &0, &  \beta\\
\end{pmatrix}
\end{equation}
with $\alpha = \beta = 1/\sqrt{2}$.
The following MW $\pi $ pulse refocuses the coherence. At the formation of the echo, the RF $\pi$ pulse on $m_I=-1/2 \leftrightarrow m_I=+1/2$ followed by the MW $\pi$ pulse transfer the coherence into the nuclear spin subspace,
\begin{equation}
\begin{pmatrix}
0, & \alpha, & -\beta, &0, &  0,  \\
\end{pmatrix}
\end{equation}
The following three RF pulses, $\pi/3$ on $m_I=-3/2 \leftrightarrow m_I=-1/2$, $\pi/3$ on $m_I=+1/2 \leftrightarrow m_I=+3/2$, and RF $-\pi$ on $m_I=-1/2 \leftrightarrow m_I=+1/2$, transform the state to
\begin{equation}
-
\begin{pmatrix}
\frac{1}{2}\alpha, & \frac{\sqrt{3}}{2}\beta, &\frac{\sqrt{3}}{2}\alpha, &\frac{1}{2}\beta , &   0 \\
\end{pmatrix}
\end{equation}
corresponding, within a global undetectable phase, to the logical qubit state $\alpha\ket{0_L}+\beta\ket{1_L}$.


\subsection{Sequence for detection of $m_I=\pm1/2$ and $m_I=\pm3/2$ coherences}

Following encoding, error application and refocusing, the state of the system is, in general,
\begin{equation}
\begin{pmatrix}
\xi, \zeta, \eta, \lambda, 0
\end{pmatrix}
\end{equation}
As explained in the main manuscript, owing to the incomplete refocussing, the experimental objective is to detect the $m_I=\pm1/2$ and $m_I=\pm3/2$ coherences via electron spin echoes.

The RF $\pi $ pulse on $m_I=-1/2 \leftrightarrow m_I=+1/2$ changes this state to
\begin{equation}
\begin{pmatrix}
\xi, -\eta, \zeta, \lambda, 0
\end{pmatrix}
\label{det-transitional}
\end{equation}

Without the conditional (green) pulses, the coherence transfer sequence of a MW $\pi$ pulse and an RF $\pi$ pulse on $m_I=-1/2 \leftrightarrow m_I=+1/2$ yields the state
\begin{equation}
\begin{pmatrix}
\xi, -\zeta, 0, \lambda, -\eta
\end{pmatrix} \: .
\end{equation}
The subsequent MW refocusing pulse generates an electron spin echo from the amplitudes $[\eta,-\zeta]$, yielding the quantities $I_{\pm\frac{1}{2} x}$ and $I_{\pm\frac{1}{2} y}$ in the main manuscript.

With the conditional (green) pulses, the state in Eqn.~\ref{det-transitional} is transformed to
\begin{equation}
\begin{pmatrix}
\eta, \xi, -\lambda, \zeta,  0
\end{pmatrix}
\end{equation}
The coherence transfer sequence of a MW $\pi$ pulse and an RF $\pi$ pulse on $m_I=-1/2 \leftrightarrow m_I=+1/2$ then yields the state
\begin{equation}
\begin{pmatrix}
\eta, \lambda, 0, \zeta, \xi
\end{pmatrix}
\end{equation}
The subsequent MW refocusing pulse generates an electron spin echo from the amplitudes $[-\xi,\lambda]$, yielding the quantities $I_{\pm\frac{3}{2} x}$ and $I_{\pm\frac{3}{2} y}$ in the main manuscript.

\section{Pulse Fidelity Estimations}

\begin{figure}[htb!]
\includegraphics[width=\columnwidth]{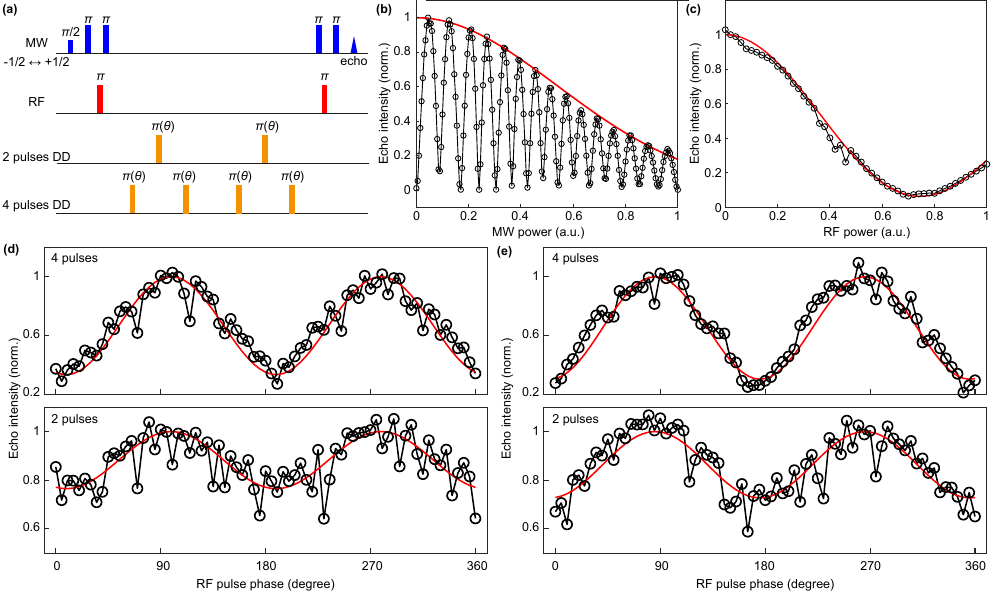}
\caption{\textbf{(a)} The pulse sequence employed to estimate the fidelity of the RF $\pi$ pulses. All the red and orange pulses are RF pulses with the same frequency. We apply either the 2-pulses DD or the 4-pulse DD pulse train. The phase ($\theta$) of the $\pi(\theta)$ pulses is swept in the experiment with the intensity of the final echo recorded as a function of $\theta$. \textbf{(b)} the Rabi oscillation of the electron spin echo with varying the amplitude of the MW pulse. \textbf{(c)} The echo intensity of a standard Davis ENDOR sequence with varying the amplitude of the RF pulse. \textbf{(d)} and \textbf{(e)} show the intensity of the final echo as a function of $\theta$ measured using the pulse sequence in \textbf{(a)}. \textbf{(d)} and \textbf{(e)} are measured with the RF frequency of 83.2~MHz ($\ket{-1/2}\otimes\ket{-3/2} \leftrightarrow \ket{-1/2}\otimes\ket{-3/2}$) and 87.4~MHz ($\ket{-1/2}\otimes\ket{-1/2} \leftrightarrow \ket{-1/2}\otimes\ket{+1/2}$), respectively. The upper panels are the data measured with the 4-pulse DD train whereas the bottom panels are measured with the 2-pulse DD train.}
\label{fig_fidelity}
\end{figure}

The fidelity of the MW and RF pulses are calibrated to evaluate the errors introduced in the spin control sequences. The dominant error in our control pulses is the rotation angle error due to the $B_1$ inhomogeneity associated with the resonator/RF coil profiles. The error for the MW pulse is estimated by observing the Rabi oscillation of the electron spin echo over a number of periods by varying the power of the initial pulse in an electron Hahn echo sequence. The result is shown in Fig.~\ref{fig_fidelity}(b). 

From these data we can obtain a rotation angle error estimate following the procedure described in Reference~\cite{Morton2005}. The procedure models the $B_1$ inhomogeneity with a Gaussian distribution with standard deviation $\sigma_{MW}$. For a nutation sequence of $n$ $\pi$-rotations (with $n\sigma_{MW}<1$), the $n$th maximum is smaller than the first by a factor $\exp{[-\sigma_\mathrm{MW}^2n^2]}$. Thus from the envelope of the nutation maxima, we can accurately extract the fidelity of a single $\pi$-rotation. The fidelity of our MW pulses estimated using this method is $99.5 \pm 0.1$\%.

We cannot use the same procedure for estimating the rotation angle error for our RF pulses, because, even at maximum power, we cannot obtain multiple $\pi$-rotations with a pulse of sufficient bandwidth; Fig.~\ref{fig_fidelity}(c) shows the nuclear spin nutation over our available RF power range. In this situation, we use a variant of an alternative approach described in Reference~\cite{Morton2005}: comparing a Carr-Purcell echo train (which accumulates rotation angle errors from successive $\pi$ pulses) with a Carr-Purcell-Meiboom-Gill echo train (for which rotation angle errors cancel out after each cycle of two $\pi$ pulses).

Unlike in Reference~\cite{Morton2005}, we cannot observe the train of nuclear spin echoes as they form, because our detection requires transferring the coherence to the electron spin for detection via electron spin echo. Instead, we observe the final echo intensity following a $n=2$ pulse or $n=4$ pulse dynamical decoupling (DD) sequence of $\pi$ pulses of phase $\theta$ with respect to the phase of the initial nuclear spin coherence. For $\theta = 0, \pi, 2\pi, ...$, this DD sequence is identical with a Carr-Purcell sequence. For $\theta = \pi/2, 3\pi/2, ...$ it is identical with a Carr-Purcell-Meiboom-Gill sequence. Thus as a function of $\theta$, we see an oscillation in the final echo amplitude, where the ratio between the minimum and maximum signal is given by $\exp{[-\sigma_\mathrm{RF}^2n^2]}$, where $\sigma_\mathrm{RF}$ is the standard deviation of the Gaussian distribution of RF $B_1$ fields across the sample. 

We performed this procedure for the two RF transitions from the $m_I = -1/2$ state. Fitting the variation of the final echo amplitude with $\theta$ yielded an estimate of the RF pulse fidelity of $93.5\pm0.1$\% for RF pulses at both frequencies.

\section{Calibration of the artificial phase gate}
A controlled artificial phase perturbation, $\widehat{Z(\theta)}$, was applied to investigate how the logical qubit evolves under a transient fluctuation of the magnetic field (Fig.~3 in the main text). This artificial phase perturbation is achieved by the generation of a transient magnetic field that is parallel to the static $B_0$ field. We first calibrate the effect of $\widehat{Z(\theta)}$ by measuring its effect on a standard nuclear coherence between two states. 

\begin{figure}[htb!]
\includegraphics[width=\columnwidth]{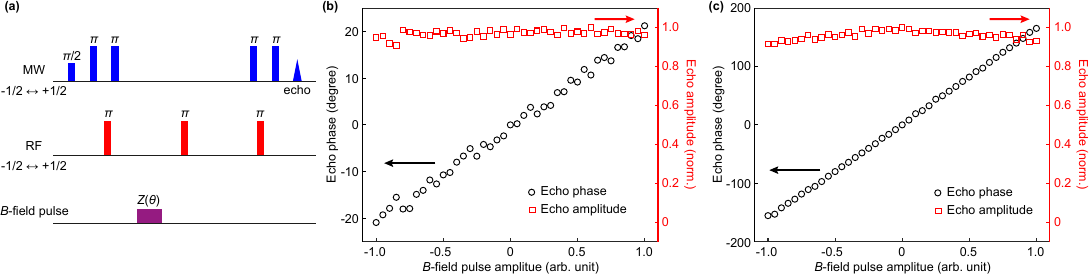}
\caption{\textbf{(a)} the pulse sequence for the calibration of the artificial phase perturbation $\widehat{Z(\theta)}$. The phase and the amplitude of the echo are measured while varying the amplitude of the $B$-field pulse with a duration of 10~$\mu$s \textbf{(b)} and 80~$\mu$s \textbf{(c)}.}
\label{fig_Z_pulse}
\end{figure}

The calibration sequence is shown in Fig.~\ref{fig_Z_pulse}(a). The spin control pulses are the same as the coherence transfer scheme described in Fig.~\ref{fig_endor}(b) with a magnetic field pulse ($B$-field pulse) applied during the first free evolution period of the nuclear spin. Such a $B$-field pulse modifies the precession frequency for the nuclear spins, leading to a phase shift in the final echo. The calibration results are shown in Fig.~\ref{fig_Z_pulse}(b) and (c), where the duration of the $B$-field pulse is fixed to 10 $\mu$s and 80 $\mu$s, respectively, while the amplitude of the pulse is varied. A linear phase dependence on the $B$-field pulse amplitude was observed for both pulse durations, with the effect of the 80 $\mu$s pulse approximately 8 times that of the 10 $\mu$s pulse. The results confirm that the phase shift is proportional to both the duration and the amplitude of the applied $B$-field pulse. Furthermore, the amplitude of the echo remains stable throughout the experiment, with a maximum 10\% modulation observed for the strongest 80 $\mu$s $B$-field pulse. This confirms that a reasonably homogeneous $B$-field pulse is applied across the sample. Any inhomogeneity in the transient magnetic field pulse will reduce the echo signal by introducing a phase distribution to the spin ensemble, similar to the ones describe in Sec.~\ref{QEC_refocus}. However, such inhomogeneity only exists during the $B$-field pulse, and therefore cannot be eliminated by the refocusing pulse.

\section{Refocusing the qudit State}
\label{QEC_refocus}

In the experiment described in this work we must consider the effect of inhomogeneous broadening across the spin ensemble. The effect of inhomogeneous broadening leads to some electron/nuclear spins detuned from the applied MW/RF radiations, resulting in an undesired phase distribution for the quantum coherence stored the spin ensemble. Such effects must be mitigated using carefully placed refocusing pulses to allow proper coherence transfers and echo detections.

\begin{figure}[htb!]
\includegraphics[width=\columnwidth]{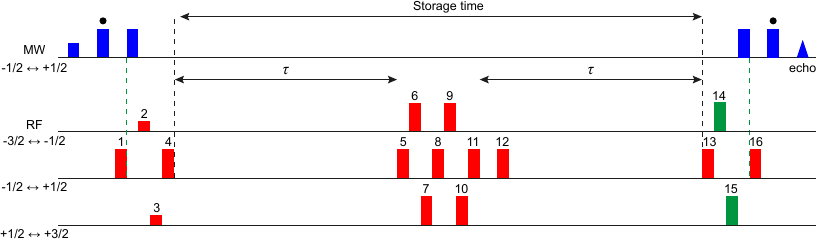}
\caption{The pulse sequence for quantum error correction with nuclear logical qubits. An electron spin coherence is generated by the very first MW pulse and then transferred to the hyperfine-coupled nuclear spin. The MW refocusing pulses are labeled with $\bullet$. The MW pulses are aligned to the RF pulse 1 and 16 as indicated by the vertical dashed lines.  All RF pulses are labeled with their information listed in Table~\ref{Tab_QEC}. The RF pulses in the dashed box serve as the refocusing operation for the nuclear logical qudit. }
\label{fig_QEC}
\end{figure}

The inhomogeneous broadening effect for the electron spins can be removed straightforwardly since the coherence is stored in the superposition of two electron spin eigenstates of $\ket{-1/2}\otimes\ket{-1/2}$ and $\ket{+1/2}\otimes\ket{-1/2}$. This allows us to eliminate the inhomogeneities in the electron spins by applying the two MW $\pi$ refocusing pulses in the encoding/decoding parts as shown in Fig.~\ref{fig_QEC}. By comparison, refocusing the nuclear qudit state is more challenging because the quantum coherence is stored in a four dimensional Hilbert space. For the quantum coherence stored in the superposition of two nuclear spins states with $\delta m_I = 1$, the inhomogeneous broadening induced phase is $\delta f_i \times \tau$, where $\delta f_i$ is the detuning of the nuclear spin Larmour frequency from the RF control radiation of $f_i$ and $\tau$ is the free evolution time of the spin. The nuclear spin transition frequencies for Mn:ZnO can be directly modulated by the nuclear Zeeman interaction, the hyperfine interaction and the nuclear quadrupole moment. The latter two couple to different transitions differently, making $\delta f_1 \neq \delta f_2 \neq \delta f_3$ and leading to three phases that cannot be eliminated using a simple refocusing pulse. Additionally, the fault tolerant memory protocol requires the coherence to be (almost) always encoded using the logical qubits, preventing us from refocusing by cycling the coherences between different superpositions periodically \cite{Vitanov2015}. 

\begin{table}[htb!]
\centering
\caption{The parameters for the RF pulses used in the quantum error correction sequence.}
\label{Tab_QEC}
\renewcommand{\arraystretch}{1}
    \begin{tabular}{ | c | c | c | c | c | c | c | c | c | c | c | c | c | c | c | c | c | c | c | c | c | c | c | c | c | } 
    \hline
     Pulse \# & 1 & 2 & 3 & 4 & 5 & 6 & 7 & 8 \\
     \hline
    Frequency & $f_2$ & $f_1$ & $f_3$ & $f_2$ & $f_2$ & $f_1$ & $f_3$ & $f_2$ \\
    \hline
    Position & 0 & $U$ & $2U$ & $3U$ & $3U + \tau$ & $4U + \tau$ & $5U + \tau$ & $6U + \tau$ \\
    \hline
    Ampl. \& phase & $+\pi$ & $+\pi/3$ & $+\pi/3$ & $-\pi$ & $+\pi$ & $\pm\pi$ & $+\pi$ & $+\pi$ \\
    \hline
    \hline
    Pulse \# & 9 & 10 & 11 & 12 & 13 & 14 & 15 & 16 \\
    \hline
    Frequency & $f_1$ & $f_3$ & $f_2$ & $f_2$ & $f_2$ & $f_1$ & $f_3$ & $f_2$\\
    \hline
    Position & $7U + \tau$ & $8U + \tau$ & $9U + \tau$ & 15U + $\tau$ & $9U + 2\tau$ & $10U + 2\tau$ & $11U + 2\tau$ & $12U + 2\tau$ \\
    \hline
    Ampl. \& phase & $\pm\pi$ & $\pm\pi$ & $-\pi$ & $+\pi$ & $+\pi$ & $+\pi$ & $+\pi$ & $+\pi$ \\
    \hline
    \end{tabular}%
\end{table}%

We design a composite refocusing sequence as shown in Fig.~\ref{fig_QEC}. Table~\ref{Tab_QEC} shows the exact position of all RF pulses involved in the fault tolerant memory sequence. In a practical experiment, one needs to take into account the finite duration of the RF pulses and the ring down behaviour of the RF coil. To compensate for these effects, we introduce small delays between consecutive RF pulses, multiples of $U$ (= 8 $\mu$s $\ll \tau$), as listed in Table~\ref{Tab_QEC}. For QEC experiemnt with artificial error (Fig.2 in main text), $\tau$ was fixed to 0.1 ms, and for experiment with varying storage time (Fig.3 in main text), $\tau$ was changed from 0.1 ms to 3 ms. Pulses 5 to 12 act effectively as a $\pi$ pulse over the four dimensional nuclear spin subspace of $m_I = -3/2, -1/2, +1/2$ and $+3/2$. Note pulses 6, 9 and 10 are alternated between $+\pi$ and $-\pi$ pulses in experiments. This alternation, together with varying the phase of the first MW coherence generation pulse, produces a four step phase cycling that reduces undesired echo contributions, similar to that described in Table~\ref{phase_cycle}. The refocusing block can be represented by the propagator
\begin{equation}
\label{pulse_pi}
\mathbf{P}_{\mathrm{refocus},\pm} = 
\begin{pmatrix}
0 & 0 & 0 & \pm1 \\ 
0 & 0 & \pm1 & 0 \\ 
0 & +1 & 0 & 0 \\ 
+1 & 0 & 0 & 0 \\ 
\end{pmatrix}
\end{equation}
where the $+$ and $-$ sign in the subscript corresponds to pulses 6, 9 and 10 being $+\pi$ or $-\pi$, respectively. Here we only present the nuclear spin part for clarity since the electron spin is always in the $m_S = -1/2$ state when the coherence is stored in the nuclear spin. 

An arbitrary nuclear superpostion state, $\ket{-1/2}\otimes\ket{\psi}$ with $\ket{\psi}$ being the nuclear spin part, undergoes a period $\tau$ of free evolution where the effect of inhomogeneous broadening has the effect:
\begin{equation}
\label{FID_tau}
\ket{\psi} = 
\begin{pmatrix}
\xi \\ \zeta \\ \eta \\ \lambda 
\end{pmatrix}
\xRightarrow[]{\text{time}\enspace\tau}
\ket{\psi(\tau)} = 
\begin{pmatrix}
\xi e^{-\mathrm{i}(\delta f_1 + \delta f_2 + \delta f_3)\tau} \\ 
\zeta e^{-\mathrm{i}(\delta f_1 + \delta f_2)\tau} \\ 
\eta e^{-\mathrm{i}\delta f_1\tau} \\ \lambda  
\end{pmatrix}
\end{equation}

$\ket{\psi(\tau)}$ is the state before the refocusing pulses are applied (immediate before pulse 5) up to an unimportant global phase. Note this effect is different from the artificial perturbation shown in the main text in that {(1)} different $m_I$ coherence gain a different phases and {(2)} the same effect will persist throughout the entire pulse sequence. The refocussing pulses then swap the coherence within the subspace, allowing the nuclear spins to evolve for another period of $\tau$ under the same inhomogeneous shifts, as 
\begin{equation}
\label{FID_2tau}
\mathbf{P}_{\mathrm{refocus},\pm}\ket{\psi(\tau)} = 
\begin{pmatrix}
\pm\lambda \\ 
\pm\eta e^{-\mathrm{i}\delta f_1\tau} \\ 
\zeta e^{-\mathrm{i}(\delta f_1 + \delta f_2)\tau} \\ 
\xi e^{-\mathrm{i}(\delta f_1 + \delta f_2 + \delta f_3)\tau} \\ 
\end{pmatrix}
\xRightarrow[]{\text{time}\enspace\tau}
\ket{\psi(2\tau)} = 
\begin{pmatrix}
\pm\lambda e^{-\mathrm{i}(\delta f_1 + \delta f_2 + \delta f_3)\tau} \\ 
\pm\eta e^{-\mathrm{i}(2\delta f_1 + \delta f_2)\tau} \\ 
\zeta e^{-\mathrm{i}(2\delta f_1 + \delta f_2)\tau} \\ 
\xi e^{-\mathrm{i}(\delta f_1 + \delta f_2 + \delta f_3)\tau} \\ 
\end{pmatrix},
\end{equation}
where $\ket{\psi(2\tau)}$ is the state ready to be decoded and transferred back to the electron spin for the read-out process. 

When the initial state is an encoded logical qubit defined by Eqn.~3 of the main manuscript, this refocusing sequence retains the quantum information using logical qubits throughout the storage time of $2\tau$. Eqn.~\ref{FID_2tau} shows that the coherence stored in the superpositions $\ket{-1/2}\otimes\ket{-3/2} \leftrightarrow \ket{-1/2}\otimes\ket{+3/2}$ and $\ket{-1/2}\otimes\ket{-1/2} \leftrightarrow \ket{-1/2}\otimes\ket{+1/2}$ are refocussed for any type of inhomogeneity. The other coherences, those in the superpositions $\ket{-1/2}\otimes\ket{-3/2} \leftrightarrow \ket{-1/2}\otimes\ket{-1/2}$ and $\ket{-1/2}\otimes\ket{+3/2} \leftrightarrow \ket{-1/2}\otimes\ket{+1/2}$ cannot be refocussed using our sequence. Nevertheless, the two recovered coherences allow us to extrapolate the information about the evolution of the encoded state using the measurement scheme described in the main text. 

\section{Overlap of logical qubit and error component with artificial error}
\label{Overlap}

\begin{figure}[htb!]
\includegraphics[width=0.5\columnwidth]{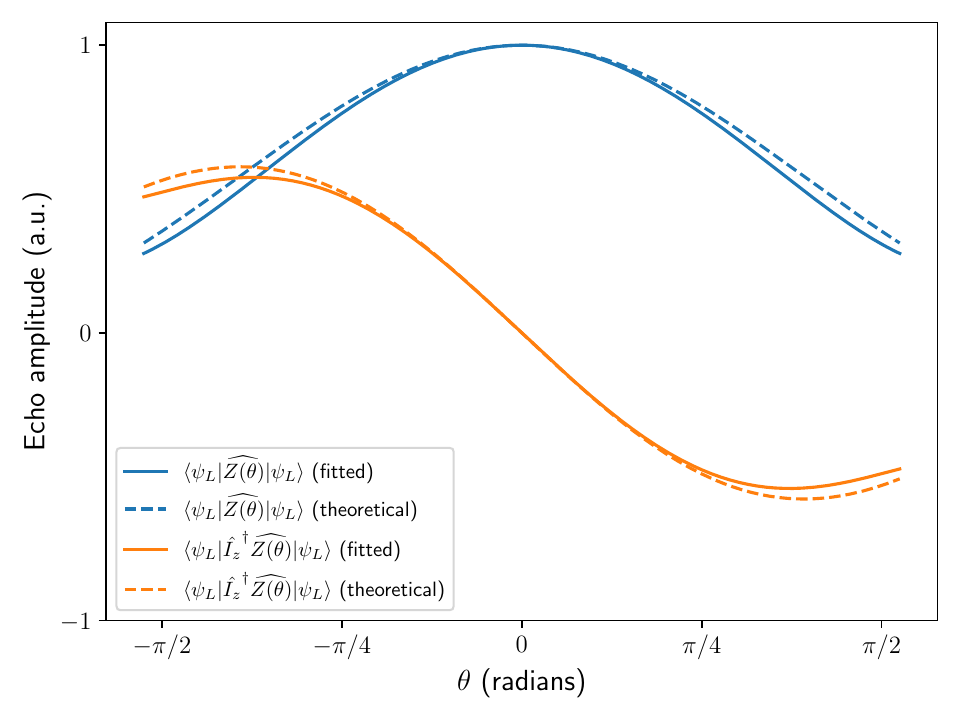}
\caption{Solid lines: Estimates of the quantities $\bra{\psi_L}  \widehat{Z(\theta)} \ket{\psi_L}$
and
$\bra{\psi_L}  \hat{I_z}^{\dagger} \widehat{Z(\theta)} \ket{\psi_L} $ extracted from the parameterised fit to the experimental data. Dashed lines: exact theroetical evolution of the same quantities.}
\label{overlaps}
\end{figure}

In Fig.~\ref{overlaps}, we present the overlap of logical qubit ($\bra{\psi_L}  \widehat{Z(\theta)} \ket{\psi_L}$) and the error component ($\bra{\psi_L}  \hat{I_z}^{\dagger} \widehat{Z(\theta)} \ket{\psi_L} $) obtained (1) by theoretical calculation (dashed lines) and (2) from the fitting of experimental data (solid lines), as a function of the artificial error $\theta$ that we applied. The measured part is constructed from the parameters $A_n$, $n = 0 \cdots 5$. As expected for small $\theta$, $\bra{\psi_L}  \widehat{Z(\theta)} \ket{\psi_L}$ is independent of $\theta$ and $\bra{\psi_L}  \hat{I_z}^{\dagger} \widehat{Z(\theta)} \ket{\psi_L}$ varies linearly with $\theta$.

This demonstrates the general point that using the parameterisation, we can explore the evolution of physically meaningful quantities that we cannot measure directly.

\section{Experimental setup}
\label{Setup}

\begin{figure}[htb!]
\includegraphics[width=1.0\columnwidth]{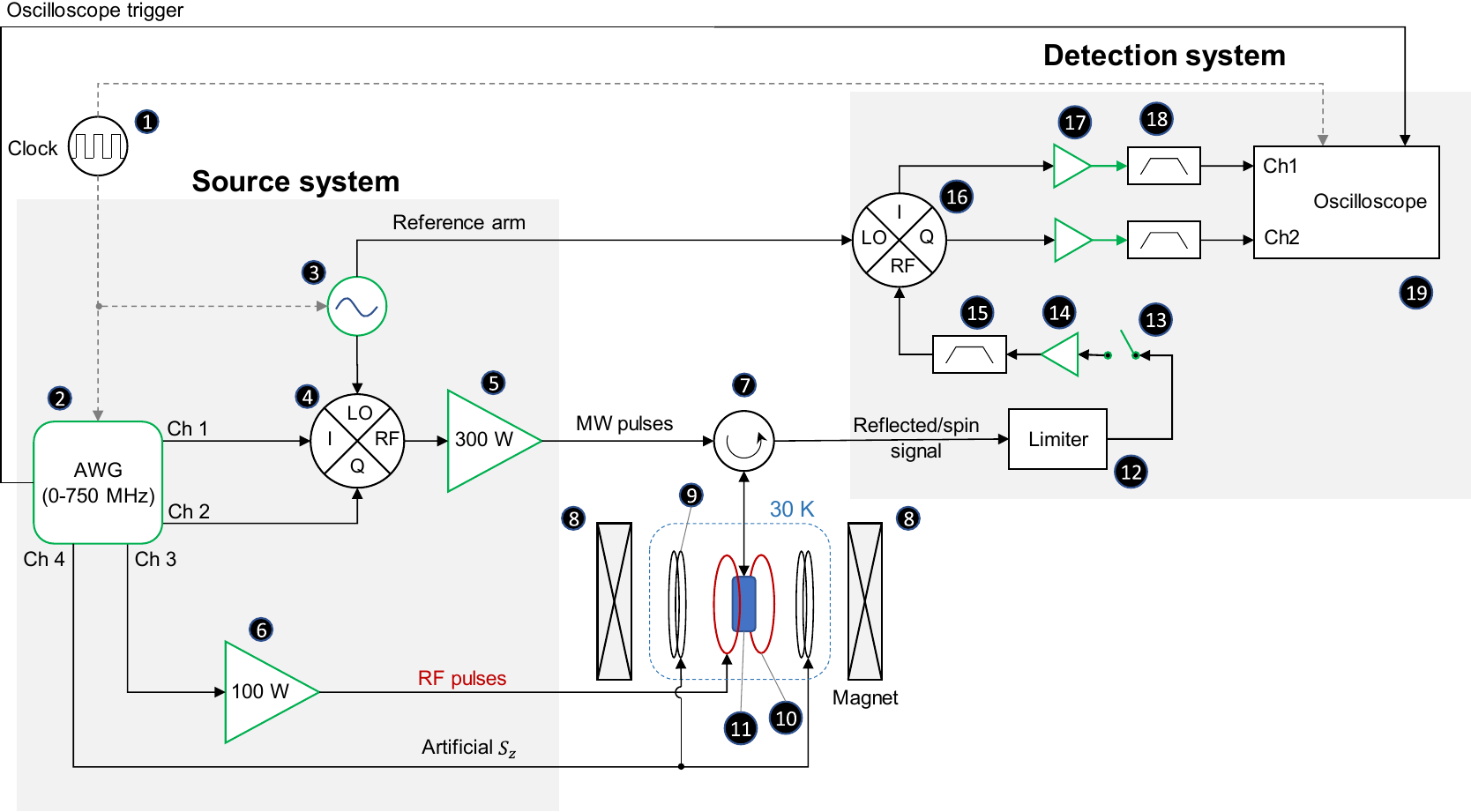}
\caption{ Schematic diagram of experimental setup in this work. }
\label{setup}
\end{figure}

Our custom-built spectrometer is sketched in Fig.~\ref{setup}. The important feature of the setup is that all time-dependent control signals are sourced from a single arbitrary waveform generator. This includes all microwave (MW) electron spin resonance pulses and radio frequency (RF) nuclear spin resonance pulses. This facilitates the implementation of the complex multifrequency coherent sequences required for this experiment.

The parts assigned with numbers are:\\ 
(1) Clock, 10MHz TTL-reference: Stanford Research Systems FS725\\
(2) Arbitrary waveform generator (AWG): Zurich instruments, HDAWG 750 MHz, 4 ports\\
(3) Microwave (MW) source: Keysight EE8257D PSG Signal Generator\\
(4) IQ-mixer for up-conversion: MLIQ-0416, 4-16 GHz\\
(5) Amplifier for MW-pulses: ERZIA ERZ-HPA-0850-0980-55, 8.5-9.8 GHz, 300 W\\
(6) Amplifier for RF-pulses: Mini-Circuits ZHL-100W-52-S+, 50-500 MHz, 100 W\\
(7) Microwave X-band circulator\\
(8) Resistive magnet: 0 - 0.97 T\\
(9-11): Bruker EN 4118X-MD4 resonator combined with an Oxford Instruments cryostat\\
\indent(9) Modulation coils\\
\indent(10) RF ENDOR coils\\
\indent(11) MW cavity, sample location\\
(12) Limiter: Narda LIM 201, 1-18 GHz, max 1 $\mu$s at 150W \\
(13) Protection switch: SP213DHTS-80, 70 dB isolation \\
(14) Cascade of low-noise MW amplifiers: Narda-MITEQ LNA-40-08001200-09-10P, and Mini-Circuits ZX60-05113LN+ \\
(15) Cavity bandpass filter: Mini-Circuits ZVBP-9750-S+, 9.5-10 GHz \\
(16) IQ-mixer for down-conversion: MLIQ-0416, 4-16 GHz \\
(17) Preamplifier: Stanford Research Systems SR445A, 350 MHz, 13dBm \\
(18) Bandpass filter: Mini-Circuits BBP-100+, 87-117 MHz \\
(19) Oscilloscope: Tektronix DPO7254 \\

During measurements, the sample is located in the microwave (MW) cavity of the Bruker MD4 resonator (11), the static magnetic field $B_0$ is generated by the resistive magnet (8), and the reference 10 MHz-signal is provided by the frequency standard (1). 

Pulse sequences are programmed in Python on an external computer and are uploaded onto the AWG (2). The  shot repetition time is determined by the longitudinal nuclear spin relaxation time $T_{1n}$ (which limits the thermalisation time) of the system under study and is encoded as a wait period in the pulse-sequence code. 

The external computer triggers the AWG (2) making it produce a required number of identical pulse-sequences (shots). At the end of each sequence, the AWG (2) sends a TTL trigger signal to the oscilloscope (19). The TTL pulse is synchronised with the RF and MW pulses. 

RF pulses are generated by the AWG (2) and get directly multiplied at (6) before they are fed to the RF coil of the Bruker MD4 resonator (10). As for MW pulses, the AWG (2) first generates the envelopes of MW pulses at 100MHz and sends them to the IQ-mixer (4), where they get mixed with the 9.6 GHz signal from the MW-source (3) and undergo up-conversion to about 9.7 GHz. After that, the MW pulses get amplified at (5) and enter the MW cavity of the MD4 resonator (11) directed by the circulator (7). A proper alignment of the MD4 resonator allows one to ensure that MW-field $B_1$, RF-field $B_2$ and the static field $B_0$ are mutually perpendicular.

The excitation MW pulses reflected from the resonator cavity and the useful echo signals generated by the spin system are directed by the circulator (7) towards the detection circuit of the spectrometer. The excitation pulses are prevented from entering and damaging the detection system by the limiter (12) and protection switch (13). The opening time of the switch defines the dead-time of the spectrometer. After the switch is closed, the signal produced by the spin system can proceed to the cascade of low-noise amplifiers (14), band-pass filter (15), and to the IQ-mixer (16) where it gets down-converted to 100 MHz using the reference arm from the MW-source (3). The use of the IQ-mixer (16) allows us to perform quadrature detection and obtain both the in-phase and quadrature components of the signal.  Both down-converted signals get amplified at (17), pass the band-pass filters (18), and reach the oscilloscope (19). The echoes resulting from separate shots of identical pulse-sequences are accumulated by the oscilloscope (19). Finally, the external computer downloads the averaged signal from the oscilloscope for further data processing.

The artificial $B_z$ fluctuation was implemented using the modulation coils of the MD-4 resonator (9). The corresponding pulses were incorporated into the pulse sequences uploaded to the AWG (2), whose output was fed to the modulation coils (9).

%